\documentstyle[12pt,epsfig]{article}
\newcommand{\be}{\begin{equation}}
\newcommand{\ee}{\end{equation}}
\newcommand{\ba}{\begin{eqnarray}}
\newcommand{\ea}{\end{eqnarray}}
\newcommand{\no}{\nonumber \\}
\newcommand{\ep}{\epsilon}
\newcommand{\gmu}{g^{\mu\nu}}
\begin{document}
\begin{titlepage}
\pagestyle{empty}
\vspace{0.5in}
%\begin{flushright}
%\today
%\end{flushright}
\vspace{0.5in}
\begin{center}
\begin{large}
{\bf{Renormalization Group Analysis of $\rho$-Meson}}\\
{\bf Properties at Finite Density}\\
\end{large}
\vskip 0.5in
Youngman Kim and Hyun Kyu Lee\\
\vskip 0.5in
{\small {\it Department of Physics, Hanyang University\\
Seoul 133-791, Korea}}
\end{center}
\vspace{1.5cm}
\begin{abstract}
We calculate the density dependence of
the $\rho$-meson mass and
coupling constant($g_{\rho NN}$)
 for $\rho$-nucleon-nucleon vertex at one loop using the
lagrangian where the $\rho$-meson is included
as a dynamical gauge boson of a hidden local symmetry. From 
the condition that thermodynamic potential should
not depend on the arbitrary energy scale,
renormalization scale, one can construct a
renormalization group equation for the thermodynamic potential and
argue that the various renormalization
group coefficients are functions of the density or temperature.
We calculate the $\beta$-function for $\rho$-nucleon-nucleon coupling constant
($g_{\rho NN}$) and $\gamma$-function for $\rho$-meson mass
($\gamma_{m_\rho}$). We found that the $\rho$-meson mass and the
coupling constant for $g_{\rho NN}$ drop as density increases
in the low energy limit.

\end{abstract}
%\hspace{.4in}PACS numbers 
\end{titlepage}

\section{Introduction}
Recently, the properties of vector mesons in dense nuclear matter, which 
give several physical consequences, have been seriously investigated
\cite{rho}\cite{friman}\cite{klingl}.
The CERES collaboration reported qualitatively the excess of dileptons 
with low invariant mass\cite{ceres}.

According to Brown/Rho(BR) scaling\cite{br}, which links  
the vector meson mass to scalar quark condensate, 
the vector meson masses will drop in nuclear matter and this gives the most
simple explanation for the observed low-mass dileptons.
In ref.\cite{kim}, the authors investigated BR scaling using 
the dilated chiral quark model. 
While the theory of BR scaling is based on the idea of quasiquarks
(partonic picture), Rapp, Chanfray and Wambach\cite{rapp}
 provide an alternative 
point of view(hadronic picture), 
which is based on the conventional many body theory, to
explain the experimental result.
In ref.\cite{rho}, the authors tried to construct a model that interpolates
the theory of ref.\cite{rapp} to the BR scaling.

Most of the works on the vector meson properties focused 
on the $\rho$-meson mass
because  it is plausible that the short-lived $\rho$-meson produced 
in heavy-ion collisions couple to photons and decay into the lepton
pairs in medium. Besides the in-medium $\rho$-meson mass, 
the coupling constants for $\rho$-meson-hadrons could 
change with the density or temperature. The change of the coupling constant 
in medium will affect the in-medium $\rho$-meson mass\cite{rho}.
So the clear understanding of the in-medium properties
(mass, coupling constants) of $\rho$-meson is important.

To describe the strongly  interacting hadronic systems  at low energy,
various kinds of effective theories of Quantum Chromodynamics(QCD) have been
suggested. As suggested by Bando, et al\cite{kugo},
the $\rho$-mesons can be identified as the dynamical gauge boson of
hidden local symmetry in the $SU(2)_L\times SU(2)_R/SU(2)_V$ nonlinear chiral
lagrangian. Besides the lowest configurations of effective interactions
in our model lagrangian, we include the effects of the vector meson-nucleon
tensor coupling. This tensor coupling is important in describing the 
short- and intermediate-range nucleon-nucleon force in the one-boson 
exchange model and play an important role in vector meson-nucleon 
scattering\cite{klingl}. The effects of tensor coupling on the vector mesons
propagating in dense matter is investigated in ref.\cite{roy}.
To study the changes of parameters in the lagrangian in medium, we
resort to the renormalization group equation.
Since the thermodynamic potential does not depend on
the renormlization scale, we can construct the
renormalization group equation for thermodynamic potential.
There are slightly different points of view\cite{ka,ma}.
In the case of renormalization group equations 
for $n$-point Green function
(proper vertex) which depend on the external momentum,
there are two typical energy scales in the system,
i.e. density(temperature) and external momentum. But in the case of
thermodynamic potential, the only typical energy scale is the
density(temperature). 
Using the standard methods in renormalization group analysis,
we can define the density dependent parameters(masses, coupling constants)
\cite{cp,moki,mc}. In ref.\cite{mc}, they showed that the density dependent
gauge coupling constant of quark gas approaches zero at high density and
argued that a phase transition between qaurk matter and hadronic matter occurs
at finite density.

The aim of this work is to study the properties of $\rho$-meson 
at finite density. We focus on the $\rho$-meson mass and 
$\rho$-nucleon-nucleon coupling constant($g_{\rho NN}$) in dense nuclear
matter.
We construct effective lagrangian following
Bando, et al.\cite{kugo} in section 2.
We discuss the renormalization group equation
for thermodynamic potential
 in section 3. We
calculate the $\beta$-function and $\gamma_{m_\rho}$ 
without and with the effects of vector meson-nucleon tensor coupling 
and solve the renormalization group equations
in section 4. In section 5, we summarize and discuss the results.
In Appendix, we present a simple example to show how to calculate 
thermodynamic potential (pressure) and to  see how the  renormalization group equation
 works which we set up
 with some dimensional coupling constants 
$\frac{\lambda}{f_\pi}$.

\section{The model lagrangian }
Following Bando, et al. \cite{kugo},
the $\rho$-meson can be identified as the dynamical gauge boson of hidden
local symmetry in the $SU(2)_L\times SU(2)_R/SU(2)_V$ nonlinear chiral
lagrangian. This means that the kinetic term of the external gauge
boson($\rho$-meson) is assumed to be generated via underlying QCD dynamics or
quantum effects at composite level.

The $[SU(2)_L\times SU(2)_R]_{global}\times[SU(2)_V]_{local}$ linear model
can be constructed in terms of two $SU(2)$-matrix valued variables,
\be
U(x)=e^{2i\pi(x)/f_\pi} =\xi_L^\dagger (x)\xi_R (x) ~~[\pi(x)\equiv \pi(x)^a
\tau^a/2]
\ee
where $\xi(x)_{R,L} =e^{i\sigma(x)/f_\pi}e^{\pm i\pi(x)/f_\pi}$.
The covariant derivative is defined by
$D_\mu\xi_L=(\partial_\mu - iV_\mu )\xi_L$ and
$V_\mu (x)$ is identified with the $\rho$-meson.

Let's introduce matter field, $\Psi(x)$, as a fundamental
representation  of $[SU(2)]_{local}$ which will be identified with
nucleon field hereafter.
Thus we can write down the lagrangian with
$[SU(2)_L\times SU(2)_R]_{global}\times[SU(2)_V]_{local}$ and with the lowest
derivatives after rescaling the gauge field, $V_\mu\rightarrow g V_\mu$,
\ba
{\it L}&=& -\frac{1}{4}(F_{\mu\nu}^{(V)})^2
  +af_\pi^2 tr(gV_\mu -\frac{\partial_\mu\xi_L\cdot\xi_L^\dagger
         + \partial_\mu\xi_R \cdot \xi_R^\dagger}{2i})^2\no
&& + f_\pi^2 tr(\frac{\partial_\mu\xi_L\cdot\xi_L^\dagger
         - \partial_\mu\xi_R \cdot \xi_R^\dagger}{2i})^2\no
&&    + \bar\Psi (x) i\gamma^\mu [\partial_\mu -ig V_\mu(x)
         ]\Psi (x)
       -m\bar\Psi(x)\Psi(x)\no
&&    +\kappa \bar\Psi(x)\gamma^\mu (\alpha_{\| \mu}(x )- g V_\mu(x) )\Psi(x)
        +\lambda\bar\Psi\gamma_5\gamma^\mu\alpha_{\mu\bot}(x )
        \Psi\label{lag1}
\ea
where
\ba
F_{\mu\nu}^{(V)} &=&\partial_\mu V_\nu -\partial_\nu V_\mu -i
[V_\mu ,V_\nu],\no
\alpha_{\mu \| , \bot}(x) &=&\frac{\partial_\mu\xi_L (x) \xi_L^\dagger (x)\pm
\partial_\mu\xi_R (x) \xi_R^\dagger (x)}{2 i}.
\ea
To quantize the theory, we introduce the gauge fixing terms 
and the ghost fields\cite{ha}.
\ba
L_{GF}=&&-\frac{1}{\alpha} tr[(\partial\cdot V)^2]+\frac{1}{2}iagf_\pi^2tr[
\partial\cdot V (\xi_L -\xi_L^\dagger+\xi_R -\xi_R^\dagger)]\no
 &&+\frac{1}{16} \alpha a^2g^2f_\pi^4
\{tr[(\xi_L -\xi_L^\dagger+\xi_R -\xi_R^\dagger)^2] \no
&&-\frac{1}{2}(tr[\xi_L -\xi_L^\dagger+\xi_R -\xi_R^\dagger])^2\}
\ea
The corresponding ghost term is
\be
L_{FP}=i tr [\bar v \{2\partial^\mu D_\mu
v+\frac{1}{2}g^2\alpha f_\pi^2a (v\xi_L+\xi_L^\dagger v + v\xi_R
+\xi_R^\dagger v )\} ]
\ee
In this work, we will choose the Landau  gauge($\alpha =0$)\cite{hara}.
We expand $\alpha_{\|\mu}(x)$, $\alpha_{\bot\mu}(x)$ and
$\xi_{L,R}$ in eq.(\ref{lag1}) to write down the interactions explicitly
in terms of fields of nucleon and pions
and also to define the relevant coupling constants and masses. We make use of
the following lowest configurations of the effective interactions.
\ba
{\it L}&=& -\frac{1}{4}(F_{\mu\nu}^{(V)})^2
      +\frac{1}{2}\partial_\mu \pi \partial^\mu \pi
   +\frac{1}{2}\partial_\mu \sigma \partial^\mu \sigma \no
&&    + g_{\rho\pi\pi }\vec V^\mu (x)\cdot (\vec\pi\times\partial_\mu\vec\pi)
       + g_{\rho\sigma\sigma} \vec V^\mu (x)\cdot
(\vec\sigma\times\partial_\mu\vec\sigma)+\frac{1}{2}m_\rho^2V_\mu^2\no
&&+ \bar\Psi (x) i\gamma^\mu [\partial_\mu -ig_{\rho NN} V_\mu(x)]\Psi (x)
       -m\bar\Psi (x)\Psi (x)\no
&&    -\frac{\kappa}{2f_\pi^2}\bar\Psi (x)\gamma^\mu(\vec\pi
     \times\partial_\mu
      \vec\pi)^a\frac{\tau^a}{2}\Psi(x)
    +\frac{\lambda}{f_\pi}\bar\Psi (x)
    \gamma_5\gamma^\mu \partial_\mu\pi\Psi (x) \no
&& +\frac{\kappa}{f_\sigma}\bar\Psi (x)\gamma^\mu \partial_\mu\sigma\Psi (x)
 -\frac{\kappa}{2f_\sigma^2}\bar\Psi (x)
  \gamma^\mu(\vec\sigma\times\partial_\mu
      \vec\sigma)^a\frac{\tau^a}{2}\Psi(x) \label{lag2}
\ea
where
\ba
f_\sigma^2&=&af_\pi^2 ,~ m_\rho^2 =ag^2f_\pi^2,~ g_\rho = agf_\pi^2,\no
g_{\rho\pi\pi} &=&\frac{1}{2}ag,~ g_{\rho\sigma\sigma} =\frac{1}{2}g ,~
g_{\rho NN}=g(1-\kappa)\label{defi}
\ea
and $\kappa$ and $\lambda$ are two arbitrary constants. 
Bando et al.\cite{kugo} and Furui et al.\cite{furi} pointed out that $\kappa$
cannot be appreciably large to guarantee
vector meson dominance and $\rho$ meson universality.
The value of $\lambda$ will be estimated in section 4.
Since the variable $\kappa$ can not be large\cite{kugo} or
could be zero\cite{furi},
we will ignore the diagrams in Fig. 1 which are proportional to $\kappa$
\footnote{One may object our approximation $\kappa\rightarrow 0$
since taking this approximation may imply that we neglect
all the interaction terms between even number of pions and nucleons and
so hidden local symmetry  with $\kappa \rightarrow 0$
is not useful if compared with the chiral lagragians
in Ref. \cite{gasser}. But we should remember
that $\rho$ meson in hidden local symmetry is nothing but even number of 
pions, $\vec V_\mu =\frac{1}{gf_\pi^2}\vec\pi\times\partial_\mu\vec\pi +...$,
though this feature is not so manifest when we add kinetic term of $\rho$-meson
to the lagrangian 
to incorporate the idea of dynamical gauge boson. 
As it is shown in Ref. \cite{kugo}, the relation,
$\vec V_\mu =\frac{1}{gf_\pi^2}\vec\pi\times\partial_\mu\vec\pi +...$, plays
very important role in
 showing the equivalence of lagrangian possessing hidden local symmetry 
 with the nonlinear chiral lagrangian. } throughout this work.
Note that Fig. 1(a) vanishes due to isospin symmetry.
\begin{figure}[hbt]
%\vskip -1.5cm
\centerline{\epsfig{file=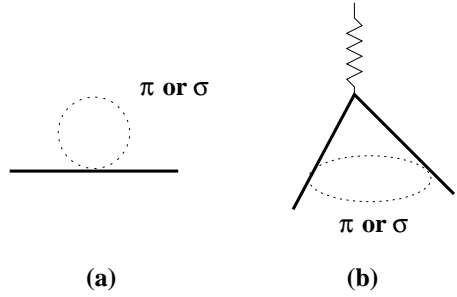,width=6cm}}
%\vskip -1.6cm
\caption{\small The diagrams which are proportional to $\kappa$. 
 Thick solid lines represent the nucleon and and wavy line is for 
$\rho$-meson. }
\end{figure}

We also include the effects of vector meson-nucleon tensor coupling.
The interaction lagrangian is given by
\be
L_{VN}=\frac{g\kappa_\rho}{2m}\bar\Psi(x)\sigma_{\mu\nu}\partial^\mu
V^\nu\Psi(x).
\ee
In reanalysis of the spin-isospin interaction, comparing with 
the spin correlation
experiments, Brown, Osnes and Rho\cite{bor} found that $\kappa_\rho=6.6$.
In the Bonn potential, $\kappa_\rho=6.1$ is used\cite{bon}. Here, we just take
$\kappa_\rho\simeq 6$. 

\section{Renormalization group equation at finite density}

At finite density, we can define the in-medium $\beta$-function
by the standard procedure\cite{cp,moki,mc}.
The condition, $\Lambda\frac{d\Omega_R}{d\Lambda}=0$,
which requires that the thermodynamic potential($\Omega_R$) or
pressure should not depend on the
renormalization point($\Lambda$), gives the following
renormalization group equation for the thermodynamic potential.
\ba
[\Lambda\frac{\partial}{\partial \Lambda} +\beta (\alpha_R)
\frac{\partial}{\partial \alpha_R} ]
\Omega_R (\alpha_R,\mu,\Lambda) =0\label{aha}
\ea
where $\beta(\alpha_R)=\Lambda\frac{\partial\alpha_R}{\partial \Lambda}$,
$\alpha_R$ is a renormalized coupling constant
and $\mu$ is a chemical potential.
It should be noted that the renormalized thermodynamic potential, $\Omega_R$,
is defined by \cite{moki,mc,bla} $
\Omega_R(g_R, m_R, T, \mu)= \Omega_B(g_B, m_B, T, \mu)-
\Omega_B(g_B, m_B, T=0, \mu=0)$.
Since  the thermodynamic potential(or pressure) 
has a mass dimension 4, it should satisfy the identity:
\ba
[\Lambda\frac{\partial}{\partial \Lambda}+
\mu\frac{\partial}{\partial \mu} ]
\Omega_R (\alpha_R,\mu,\Lambda) =4\Omega_R (\alpha_R,\mu,\Lambda)\label{id1}
\ea
By combining eq.(\ref{aha}) and eq.(\ref{id1}), we can obtain
\ba
[\mu\frac{\partial}{\partial \mu} -\beta (\alpha_R)
\frac{\partial}{\partial \alpha_R} -4]
\Omega_R (\alpha_R,\mu,\Lambda) =0.\label{rge2}
\ea
Then, we can find the general solution of $\Omega_R$ using the ansatz
\ba
\Omega_R (\alpha_R,\mu,\Lambda) =\mu^4\bar\Omega_R
 (\alpha(\mu),\mu,\Lambda)\label{sol1}
\ea
which satisfies
\ba
[\mu\frac{\partial}{\partial \mu} -\beta (\alpha(\mu))
\frac{\partial}{\partial\alpha(\mu)} ]
\bar\Omega_R (\alpha(\mu),\mu,\Lambda) =0\label{sol2}
\ea
where,
\ba
\mu\frac{\partial\alpha(\mu)}{\partial\mu} =\beta (\alpha(\mu)).\label{be5}
\ea
Eq.(\ref{be5}) states that the dependence of the  coupling constant on the
renormalization scale transmuted into the dependence 
on the density(chemical potential, $\mu$).
This discussion makes clear that $\mu$ plays the identical role in
dense matter as the spacelike external momentum does in free space
 free space renormalization\cite{moki}.

We can demonstrate it explicitly. 
Let's consider the thermodynamic potential for massless QED
at finite density as a simple example.
The thermodynamic potential(pressure)
 up to order of $\alpha^2$ can be written\cite{mc}
\ba
\Omega_R^{QED}
=\frac{1}{3\pi^2}\frac{1}{4}\mu^4[1-\frac{3}{2}\frac{\alpha(\Lambda)}{\pi}
-\frac{3}{2}(\frac{\alpha(\Lambda)}{\pi} )^2
ln(\frac{\alpha(\Lambda)}{\pi})
-\frac{1}{2}(\frac{\alpha(\Lambda)}{\pi} )^2
ln(\frac{\mu^2}{\Lambda^2})]\label{th1}
\ea
where $\alpha=\frac{e^2}{4\pi}$.
Using the ansatz of eq.(\ref{sol1}) which satisfies eq.(\ref{sol2}),
$\beta(\alpha(\mu))$ can be explicitly calculated
\ba
\mu\frac{d\alpha}{\partial\mu} =\beta (\alpha)
=\frac{2\alpha^2}{3\pi}\nonumber
\ea
which is the same as that of QED $\beta$-function in free space($\rho=0$).
Therefore we can see that  the $\beta$-function in medium has same the form
as that in free space.

We can do the same analysis with mass. The renormalization group
equation for the thermodynamic potential is given by
\ba
[\Lambda\frac{\partial}{\partial \Lambda} +\beta (\alpha_R)
\frac{\partial}{\partial \alpha_R} - m_R \gamma_m(\alpha_R)\frac{\partial}
{\partial m_R}]
\Omega_R (\alpha_R,m_R,\mu,\Lambda) =0,
\ea
where
$\beta(\alpha_R) =\Lambda\frac{\partial}{\partial \Lambda} \alpha_R$,
$\gamma_m = \Lambda\frac{\partial}{\partial \Lambda} \ln Z_m $ and
$m_B=Z_m m_R$ which are defined in free space($\rho =0$).
We can obtain the renormalization group equation at finite density
using the similar procedure used in deriving eq.(\ref{rge2}) from 
eq.(\ref{aha}) and eq.(\ref{id1}),
\ba
&&[\mu\frac{\partial}{\partial \mu} -\beta (\alpha_R)
\frac{\partial}{\partial \alpha_R}
+ (1+ \gamma_m(\alpha_R) )m_R\frac{\partial}
{\partial m_R}-4]
\Omega_R (\alpha_R,m_R,\mu,\Lambda) =0.\nonumber
\ea
By defining
\ba
\mu\frac{d}{d \mu} \alpha(\mu) &=&\beta(\alpha(\mu))\\
\mu\frac{d}{d \mu}m(\mu)&=& -[1+ \gamma_m(\alpha(\mu)) ]
m(\mu)\label{mgamma}
\ea
we can obtain the general solution of the equation,
\ba
\Omega_R (\alpha_R,m_R,\mu,\Lambda)=\mu^4 \bar
\Omega_R (\alpha(\mu),m(\mu),\mu,\Lambda).
\nonumber
\ea
Here we also see that the $\gamma_m$ in medium has same form
with that in free space. 
Then, we can study the density dependence of mass
using eq.(\ref{mgamma}).

We have several coupling constants and two masses in eq.(\ref{lag2}). Now
we extend the formalism to incorporate such couplings and masses.
Since we are interested in the density(scale) dependence of $\rho$-meson mass
and $g_{\rho NN}$ at one-loop, let us 
construct the renormalization group equation  
with the parameters which is essential to our analysis in section 4.
 We define the $g_i$ with $i=1,2,3$ as 
$ g_1=g,~g_2=g_{\rho NN},~g_3=g_{\rho \pi\pi}$ and $f_\lambda =
\frac{\lambda}{f_\pi}$.  

Then the renormalization group equation for the thermodynamic potential is
given by
\ba
&&[\Lambda\frac{\partial}{\partial \Lambda} +\sum_i\beta_i (g_{i},mf_\lambda)
\frac{\partial}{\partial g_i} +\beta_{f_\lambda} (g_i,mf_\lambda)
\frac{\partial}{\partial f_\lambda}
- m \gamma_m(g_i, mf_\lambda)\frac{\partial}{\partial m}\no
&&-m_\rho \gamma_{m_\rho}(g_i, mf_\lambda)\frac{\partial}
{\partial m_\rho}]
\Omega_R (g_i, mf_\lambda , m, m_\rho,\mu, \Lambda) =0\label{rgt1}
\ea
where we use $mf_\lambda$ instead of $f_\lambda$ 
because the coupling $f_\lambda$
carries a derivative(mass parameter) which is
$m$ in our one loop calculations and omit the subscript $R$ from
the renormalized masses and coupling constants for simplicity.

To get the identity like eq.(\ref{id1}), we find the following form for
the thermodynamic potential based on naive dimension counting,
\be
\Omega_R (g_i, mf_\lambda , m, m_\rho,\mu, \Lambda)=
\Lambda^4 \bar{\Omega}_R (g_i, mf_\lambda , \frac{m}{\Lambda}, 
\frac{m_\rho}{\Lambda},\frac{\mu}{\Lambda})\label{id2-1}
\ee
where $\bar{\Omega}_R$ is dimensionless and satisfies
\be
[\Lambda\frac{\partial}{\partial \Lambda}+
\mu\frac{\partial}{\partial \mu}+m\frac{\partial}{\partial m} +
m_\rho\frac{\partial}{\partial m_\rho}
-f_\lambda\frac{\partial}{\partial f_\lambda}]
\bar{\Omega}_R (g_i, mf_\lambda , \frac{m}{\Lambda}, 
\frac{m_\rho}{\Lambda},\frac{\mu}{\Lambda})=0.\label{id2-2}
\ee
In eq.(\ref{id2-2}), the factor $f_\lambda\frac{\partial}{\partial f_\lambda}$
is included because the partial derivative $\frac{\partial}{\partial m}$
operates not only  the argument $\frac{m}{\Lambda}$ but also 
 $mf_\lambda$ in ${\bar\Omega}_R$.
>From eq.(\ref{id2-1}) and eq.(\ref{id2-2}), we  
get the identity like eq. (\ref{id1})
\ba
[\Lambda\frac{\partial}{\partial \Lambda}+
\mu\frac{\partial}{\partial \mu}+m\frac{\partial}{\partial m} +
m_\rho\frac{\partial}{\partial m_\rho}
-f_\lambda\frac{\partial}{\partial f_\lambda}-4]
\Omega_R (g_i,m, m_\rho, mf_\lambda,\mu,\Lambda)=0\label{id2}.
\ea
>From eq.(\ref{rgt1}) and eq.(\ref{id2}), 
the renormalization group equation can be written
\ba
&&[\mu\frac{\partial}{\partial \mu} -\sum_i\beta_i (g_{i},mf_\lambda)
\frac{\partial}{\partial g_i} -(\beta_{f_\lambda} (g_i,mf_\lambda)
+f_\lambda)\frac{\partial}{\partial f_\lambda}
+m (1+\gamma_m(g_i, mf_\lambda))\frac{\partial}{\partial m}\no
&&+m_\rho (1+\gamma_{m_\rho}(g_i, mf_\lambda))\frac{\partial}
{\partial m_\rho}-4]
\Omega_R (g_i, mf_\lambda, m, m_\rho,\mu, \Lambda) =0\label{rgfft}
\ea
To solve this equation, we introduce density dependent(effective or running) 
coupling constants and masses.
\ba
\mu\frac{d}{d \mu} g_i(\mu) &=&\beta_i(g_i(\mu),m(\mu)f_\lambda(\mu))\no
\mu\frac{d}{d \mu}m(\mu)&=& -[1+ \gamma_m(g_i(\mu), m(\mu)f_\lambda(\mu))
 ]m(\mu)\no
\mu\frac{d}{d \mu}m_\rho(\mu)&=& -[1+ \gamma_{m_\rho}(g_i(\mu),  
m(\mu)f_\lambda(\mu)) ]m_\rho(\mu)\no
\mu\frac{d}{d \mu}f_\lambda(\mu)&=& \beta_\lambda(g_i(\mu), 
m(\mu)f_\lambda(\mu)) +f_\lambda(\mu)
\ea
Then eq.(\ref{rgfft}) has the solution
\be
\Omega_R (g_i, mf_\lambda , m, m_\rho,\mu, \Lambda)=\mu^4
{\bar \Omega}_R 
(g_i(\mu), m(\mu)f_\lambda(\mu), m(\mu), m_\rho(\mu),\mu, \Lambda).
\ee
Here we conclude that the density dependent $\beta$-functon 
for $g_{\rho NN}(g_2)$ and $\gamma_{m_\rho}$ have the same form with
those defined in free space.

\section{Calculations of $\beta$-function and $\gamma_m$}

The $\beta$-function for
$\rho$-nucleon-nucleon vertex($g_{\rho NN}$) can be calculated using  the
following equations
\ba
\beta(g_{\rho NN})&=&\Lambda\frac{\partial g_{\rho NN}}{\partial\Lambda},\no
g_{\rho NN}^B&=&g_{\rho NN}\Lambda^\ep Z_1 Z_2^{-1}Z_3^{-1/2}
\ea
defined in {\it free space}.
Here $Z_1$ is the renormalization constant for $g_{\rho NN}$,
$Z_2$ that for the nucleon wave function defined by
$\Psi_B^{\mu}= \sqrt{Z_2}\Psi^\mu$,
 and $Z_3$ that for the $\rho$-meson
wave function defined by $V_B^{\mu}=\sqrt{Z_3}V^\mu$.
The $\ep$ is defined by $d=4-2\ep$ with the $d$ denoting spacetime dimension.
 
\begin{figure}
\vskip -1.5cm
\centerline{\epsfig{file=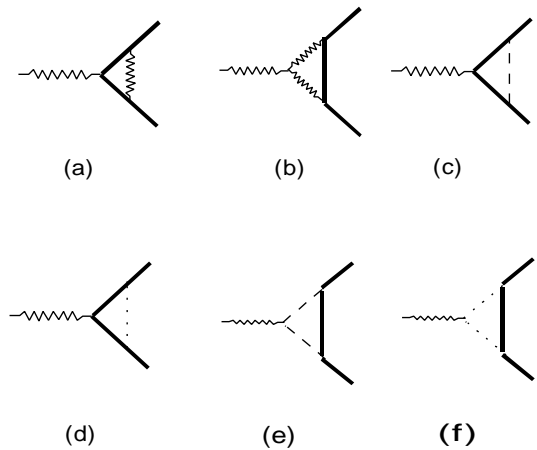,width=15cm}}
\vskip -1.6cm
\caption{\small Vertex diagrams for $g_{\rho NN}$ to one loop order. 
 Thick solid lines represent the nucleon and wavy lines are for 
$\rho$-meson. Dashed lines represent the $\sigma$ and dotted lines
are for pions respectively. }\label{fig1}
\end{figure}

We begin by calculating the $\rho$-nucleon-nucleon vertex 
function to obtain renormalization constant $Z_1$.
In Landau gauge, the contribution of the diagram in Fig. \ref{fig1}(a) 
is finite.

We evaluate the $\rho$-meson contribution in Fig. \ref{fig1}(b).
 It is calculated to be
\ba
[ig_{\rho NN}\Lambda_\mu^a]_{2b}
= g_{\rho NN}^2 g \frac{3i}{32\pi^2}\gamma_\mu T_a\frac{1}{\ep}
 +~finite~terms.
\ea
In this calculation hereafter, we take the low energy limit in which
four momentum of external $\rho$-meson is zero($k^\mu=0$)\cite{ha}.
We calculate the contribution of pion field in Fig. \ref{fig1}(f),
\ba
[ig_{\rho NN}\Lambda_\mu^a]_{2f}
= ig_{\rho\pi\pi} (\frac{\lambda}{f_\pi})^2\frac{1}{8\pi^2}
m^2 \gamma_\mu T^a\frac{1}{\ep} +~finite~terms.
\ea
The contribution of the other diagrams in Fig. \ref{fig1}
 are found to be finite. 
Hence these diagrams are not relevant as far as the infinite renormalizations
 are concerned.
Then, the renormalization constant $Z_1$ required to cancel the divergences
in counter-terms is given by
\ba
Z_1& =& 1 - [\frac{3}{32 \pi^2}g g_{\rho NN}
+\frac{1}{8\pi^2}\frac{g_{\rho\pi\pi}}{g_{\rho NN}}
(\frac{\lambda}{f_\pi})^2 m^2 ]\frac{1}{\ep}.\label{z1}
\ea

\begin{figure}
\centerline{\epsfig{file=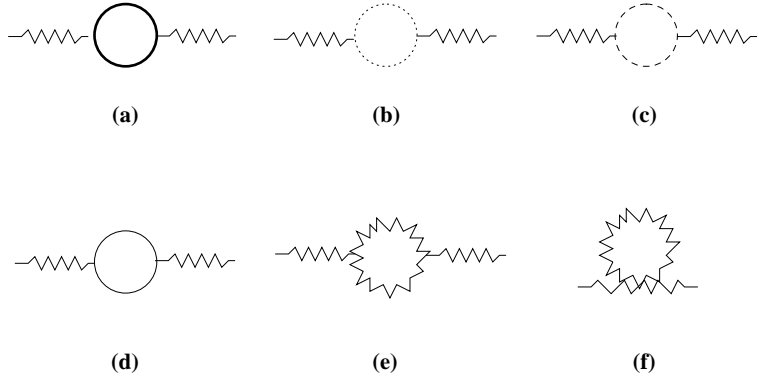,width=10cm}}
\caption{\small Self-energy diagrams for $\rho$-meson.
The thin solid lines represent the ghost.}\label{fig2}
\end{figure}

Next we turn to $\rho$-meson self-energy.
The nucleon loop contributions in Fig. \ref{fig2}
are proportional to $k^\mu k^\nu -k^2 \gmu$
 and go to zero in the low energy limit($k^\mu=0$).

The contributions of ghost loop, pion loop and
$\sigma$-loop in Fig.  \ref{fig2} are zero in dimensional 
regularization scheme since
there are no parameters which have mass dimension.
The contributions of $\rho$-meson loop and $\rho$-meson
tadpole in Fig.  \ref{fig2} does not renormalize 
the $\rho$-meson wave function
but renormalize $\rho$-meson mass in the low energy limit.
Therefore, we can take $Z_3=1$.

\begin{figure}[htb]
\vskip -10cm
\centerline{\epsfig{file=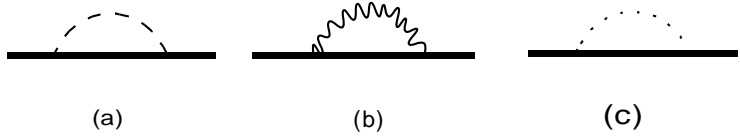,width=23cm}}
\vskip -2.1cm
\caption{\small Self-energy diagrams for nucleon }\label{fig3}
\end{figure}

Finally, we calculate nucleon self-energy diagrams.
The pion contribution is given by
\ba
[\Sigma_\pi(\not\! p)]_{4c}
=-(\frac{\lambda}{f_\pi})^2\frac{1}{16\pi^2}\frac{3}{4}
[\frac{1}{2}(p^2-m^2)\not\! p -m^2(\not\! p+m)]
\frac{1}{\ep}~+~finite~terms
\ea
and that of $\sigma$-meson in Fig. \ref{fig3} is given by
\ba
[\Sigma_\sigma(\not\! p)]_{4a}
=-(\frac{\kappa}{f_\sigma})^2\frac{1}{16\pi^2}\frac{3}{4}
[\frac{1}{2}(p^2-m^2)\not\! p -m^2(\not\! p-m)]
\frac{1}{\ep}~+~finite~terms.
\ea
In the Landau gauge,
the $\rho$-meson contributions in Fig.  \ref{fig3}
do not give wave function renormalization for nucleon fields.
Then the corresponding renormalization constant $Z_2$ is given by
\ba
Z_2 &=&1+\frac{\partial\Sigma_\pi(\not\! p)}{\partial\not\! p}|_{\not p
 =m}+
\frac{\partial\Sigma_\sigma(\not\! p)}{\partial\not\! p}|_{\not p
=m} =1.\label{z2}
\ea
>From eq.(\ref{z1}),
\ba
g_{\rho NN}^B&=&g_{\rho NN }\Lambda^\ep Z_1\no
&=&g_{\rho NN}\Lambda^\ep[
1 - (~\frac{3}{32 \pi^2}g g_{\rho NN}
+\frac{1}{8\pi^2}\frac{g_{\rho\pi\pi}}{g_{\rho NN}}
(\frac{\lambda}{f_\pi})^2 m^2 )\frac{1}{\ep}].
\ea
Then, the $\beta$-function for $g_{\rho NN}$ reads
\ba
\beta =-\frac{3 g g_{\rho NN}^2}{16\pi^2}
-\frac{1}{4\pi^2}g_{\rho\pi\pi}(\frac{\lambda}{f_\pi})^2 m^2.\label{rgef}
\ea
We can estimate the value of $\lambda$ by studying pion-nucleon interaction.
In the chiral lagrangian, pion-nucleon interaction is described by
$\frac{g_A}{f_\pi}\bar\psi\gamma_5\not\!\partial\pi\psi$\cite{gasser}
and that in our 
lagrangian is given by 
$\frac{\lambda}{f_\pi}\bar\psi\gamma_5\not\!\partial\pi\psi$. 
Then, we can conclude that $\lambda\simeq g_A$. So the value of $\lambda$
should be around $1$.
After integrating the above renormalization group equation, we obtain the
density dependent coupling constant($g_{\rho NN}$) through the arguments
given in section 3. However the procedure may not be straightforward 
 because there are many parameters (mass, coupling constants) 
which depend on the renormalization scale(density) in eq.(\ref{rgef}), 
although we can conclude from the form of the $\beta$-function 
that $g_{\rho NN}$ drops as density increases.

\begin{figure}[htb]
\centerline{\epsfig{file=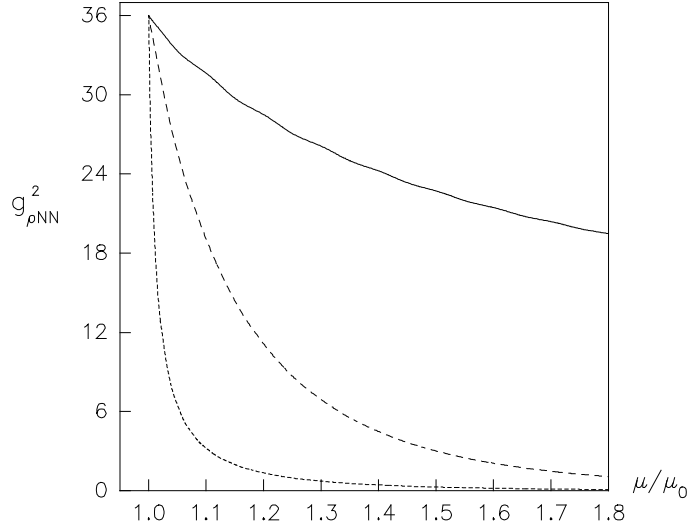,width=9cm}}
\caption{\small The density(chemical potential) dependence of $g_{\rho NN}^2$.
 Here we take $\mu_0=m$. 
The solid line is for eq.(\ref{re}) 
and the long-dashed one is for eq.(\ref{re1}). 
The short-dahsed line represents eq.(\ref{re2}).}\label{result}
\end{figure}

Consider a simplest case, for example, without pion contribution
(second term in eq.(\ref{rgef})) 
and with $\kappa =0$(where $g_{\rho NN}$ is equal to $g$).
The $\beta$-function can be written as
\ba
\beta =-\frac{3 g^3}{16\pi^2}.
\ea
Then, we can easily integrate out the renormalization group equation
to obtain
\ba
g^2(\frac{\mu}{\mu_0})=\frac{g_0^2}{1 +0.04 g_0^2\ln\frac{\mu}{\mu_0}}\label{re}
\ea
where $\mu_0$ is a reference chemical potential and
$g_0^2=g^2(\frac{\mu}{\mu_0})|_{\mu=\mu_o}$.
>From eq.(\ref{re}), 
we can easily expect that $g^2(\mu)$(or $g^2_{\rho NN}$)drops {\it slowly}
as density increases without pion contribution as depicted
in Fig. \ref{result}. Our result is qualitatively agree with the result of 
 ref.\cite{rhop}. The density dependent coupling constant $g^2$
is used in ref.\cite{song}.
As we can see in eq.(\ref{rgef}), the pion contributions
 play an important role in
decreasing $g_{\rho NN}$ in dense nuclear matter. 
To see this more explicitly, let us solve eq.(\ref{rgef}) in the
case of $\kappa=0$ and $a=2$.
In this case, the $\beta$-function is given by
\be
\beta =-\frac{3 g^3}{16\pi^2}
-\frac{1}{4\pi^2}g\frac{\lambda^2}{f_\pi^2}m^2 .\label{rgs}
\ee
To solve eq.(\ref{rgs}), we should also solve
the renormalization group equations for $\frac{\lambda}{f_\pi}$ and $m$ and 
deal with the coupled renormalization group equations. 
But here we assume as a first approximation that the $\lambda$ and $a$ 
remains constant in dense matter
and take $\lambda=1$. In ref.\cite{hara}, it is shown that the parameter
 $a$ does not change very much against temperature.
We also assume the relation,  
$\frac{m}{f_\pi}\simeq\frac{m^*}{f_\pi^*}\simeq 10$,  
to parameterize the density dependence of the
 $\frac{m}{f_\pi}$ economically 
\footnote{Here $m^*$ means the in-medium nucleon mass.
If we use the scaling law proposed in ref.\cite{rho88}, 
$\frac{m^*}{m}\simeq\sqrt{\frac{g_A^*}{g_A}}\frac{f_\pi^*}{f_\pi}$, 
we can derive the relation trivially. Since the $\lambda$ corresponds to
$g_A$ and we assume that $\lambda (g_A)$ remains constant in dense matter,
the scaling law becomes $\frac{m^*}{m}\simeq\frac{f_\pi^*}{f_\pi}$\cite{br}, 
which gives the relation $\frac{m}{f_\pi}\simeq\frac{m^*}{f_\pi^*}$.
But we are not sure that we can naively apply  the scaling law
to our $\beta$-function.}.
With these assumptions, we get the solution of eq.(\ref{rgs}),
\be
g^2(\frac{\mu}{\mu_0})=\frac{2.8 g_0^2}{-0.02 g_0^2+(2.8+0.02 g_0^2)
(\frac{\mu}{\mu_0})^{5.6} }.\label{re1}
\ee
The results are depicted in Fig. \ref{result} where we take $\mu_0=m$
 for numerical purpose.
If we make use of a simple relation, 
$\rho=\frac{2}{3\pi^2}(\mu^2 - m^2)^{3/2}$, 
we find that $\mu=\mu_0=m$ corresponds to $\rho=0$ and $\frac{\mu}{m}
\simeq 1.14$ for $\rho=\rho_0 (normal ~nuclear~
matter~density)$. Of course, the simple 
relation is no longer valid at high density. As it is mentioned, 
without the pion contribution, the $g_{\rho NN}$ drops slowly
with density(solid line). If we include the pion contribution,
we can see that the $g_{\rho NN}$ drops much faster with 
density(long-dashed line) in Fig. \ref{result}.

Let's consider the  renormalization of $\rho$-meson mass.
The $\rho$-meson loop in Fig. \ref{fig2} gives
\be
[\Pi_{\mu\nu}]_{3e}=\frac{g^2}{16\pi^2}m_\rho^2 6 \gmu\frac{1}{\ep}
\ee
and $\rho$-meson tadpole in Fig. \ref{fig2} gives
\be
[\Pi_{\mu\nu}]_{3f}=-\frac{g^2}{16\pi^2}m_\rho^2\frac{9}{2}\gmu\frac{1}{\ep}
\ee
The corresponding renormalization constant $Z_{m_\rho}$
, which is defined as $m_\rho^B= Z_{m_\rho}m_\rho^R$, is given by
\ba
Z_{m_\rho} =1-\frac{3}{4}\frac{g^2}{(4\pi)^2}\frac{1}{\ep}.
\ea
Then, $\gamma_{m_\rho}$ is calculated to be
\ba
\gamma_{m_\rho} =  \Lambda\frac{\partial}{\partial \Lambda} \ln Z_{m_\rho} =
\frac{3}{2}\frac{g^2}{(4\pi)^2}.
\ea
We can obtain density dependent $\rho$-meson mass
through renormalization group argument discussed
in the section 3,
\ba
\mu\frac{d}{d \mu}m_\rho(\mu)&=&
-(1+\frac{1}{2}\frac{3 g^2}{(4\pi)^2}) m_\rho(\mu)\label{rg22}\\
&\sim& -1.37m_\rho(\mu)\label{rgmr}
\ea
for $a=2$. Neglecting the density dependence of the gauge coupling
constant $g$, we get
\be
m_\rho(\mu) = (\frac{\mu_0}{\mu})^{1.37} m_\rho(\mu_0),
\ee
which shows that the $\rho$-meson mass\footnote{Our mass is not a pole mass
defined by the zero of an inverse propagator 
but a running mass parameter in thermodynamic(or effective) potential. One
can find the formal relation between running mass and pole mass
in ref.\cite{qu}.}
 should drop as density increases
as well as $\rho$-nucleon coupling constant in the low energy limit.

Let's consider the effects of vector meson-nucleon tensor coupling.
The interaction lagrangian is given by
\be
L_{VN}=\frac{g\kappa_\rho}{2m}\bar\Psi(x)\sigma_{\mu\nu}\partial^\mu
V^\nu\Psi(x)
\ee
where $\kappa_\rho\simeq 6$.
We expect that the contributions from the tensor coupling may change
the density dependence of $g_{\rho N N}$ substantially 
because of its anomalously
large coupling constant($g\kappa_\rho\sim 36$).

The modification due to tensor coupling in the Fig. \ref{fig1}(a) is given by
\be
[ig_{\rho NN}\Lambda_\mu^a]_{2a} =i\frac{1}{4}g_{\rho NN}T^a
[\frac{3g^2\kappa_\rho^2}{128\pi^2}\frac{m_\rho^2}{m^2}
-\frac{3gg_{\rho NN}\kappa_\rho}{32\pi^2}]\frac{1}{\ep}
\ee
and that in Fig. \ref{fig1}(b) is given by
\be 
[ig_{\rho NN}\Lambda_\mu^a]_{2b} =\frac{i}{16\pi^2}g T^a
[\frac{g^2\kappa_\rho^2}{8m^2}(10m^2+ 6m_\rho^2) +3gg_{\rho NN}\kappa_\rho]
\frac{1}{\ep}
\ee

>From these results and eq.(\ref{z1}), we find 
\ba
Z_1 = 1 &-& [\frac{3}{32 \pi^2}g g_{\rho NN}
+\frac{1}{8\pi^2}\frac{g_{\rho\pi\pi}}{g_{\rho NN}}
(\frac{\lambda}{f_\pi})^2 m^2 \no
&&+\frac{1}{4}
(\frac{3g^2\kappa_\rho^2}{128\pi^2}\frac{m_\rho^2}{m^2}
-\frac{3gg_{\rho NN}\kappa_\rho}{32\pi^2})\no
&&+\frac{1}{16\pi^2}\frac{g}{g_{\rho NN}}
(\frac{g^2\kappa_\rho^2}{8m^2}(10m^2+ 6m_\rho^2) +3gg_{\rho NN}\kappa_\rho)
]\frac{1}{\ep}.\label{mz1}
\ea

Since the vector meson-nucleon tensor coupling 
is proportional to the momentum of $\rho$-meson,
vector meson-nucleon tensor coupling does 
not play any role in the Fig. \ref{fig2}(a) in the low energy limit.

Consider the nucleon self-energy.
As shown in the section 4.1, there are no wave function renormalizations
for nucleon field.
But with vector meson-nucleon tensor coupling, the Fig. \ref{fig3}(b) 
gives the   
wave function renormalization for nucleon field.
The value of the Fig. \ref{fig3}(b) gets the following
additional contributions from the tensor coupling,
\ba
[\Sigma_{\rho}^{VN}(\not p)]_{4b}&=&
-\frac{3}{64\pi^2}[~\frac{g^2\kappa_\rho^2}{16m^2}(4p^2\not\! p-12m^2\not\! p
-6m_\rho^2\not\! p -12m^2 -12mm_\rho^2)\no
& &-\frac{gg_{\rho NN}\kappa_\rho}{4m}(-6p^2 +6m\not\! p +12m^2+12m_\rho^2)~]
\frac{1}{\ep}.
\ea
>From this we get
\ba
Z_2&=&1+\frac{\partial \Sigma_{\rho}^{VN}}{\partial \not\! p}|_{\not p=m}\no
&=&1+ \frac{3}{64\pi^2}[\frac{3g^2\kappa_\rho^2}{8}
\frac{m_\rho^2}{m^2} -\frac{3}{2}gg_{\rho NN}\kappa_\rho]\frac{1}{\ep}.
\ea
As a first approximation, assuming that 
 renormalization scale(density) dependence of $\kappa_\rho$ 
can be negligible, we get the modified $\beta$-function for $g_{\rho NN}$ 
\be
\beta=-2c_1 gg_{\rho NN}^2-c_2g^2g_{\rho NN}-2c_3g_{\rho\pi\pi}
(\frac{\lambda}{f_\pi})^2
-c_2g^2g_{\rho NN}-2c_4g^3 \label{rgff}
\ee
where $c_i$ are defined by
\ba
c_1&=&\frac{3}{32\pi^2}-\frac{1}{4}\frac{3\kappa_\rho}{32\pi^2}
-\frac{3}{64\pi^2}\frac{3}{2}\kappa_\rho ,\no
c_2&=&\frac{1}{4}\frac{3\kappa_\rho^2}{128\pi^2}\frac{m_\rho^2}{m^2}
+\frac{3\kappa_\rho}{16\pi^2} +\frac{3\kappa_\rho^2}{64\pi^2} \frac{3}{8}
\frac{m_\rho^2}{m^2},\no
c_3&=&\frac{1}{8\pi^2}\frac{m^2}{f_\pi^2},\no
c_4&=&\frac{1}{16\pi^2}\frac{\kappa_\rho^2}{8}(10+6\frac{m_\rho^2}{m^2}).
\ea
The renormalization group equation eq.(\ref{rgff})
can be solved with the $\kappa_\rho=6$, $\kappa=0$ and
$a=2$ which are assumed to be independent of density. The result is given by
\be
g^2(\frac{\mu}{\mu_0})=\frac{2.8g_0^2}{-1.04 g_0^2+(2.8+1.04 g_0^2)
(\frac{\mu}{\mu_0})^{5.6} }.\label{re2}
\ee
In Fig. \ref{result}, we can see that $g_{\rho NN}^2(\mu)$ drops much faster 
with the effects of tensor coupling than that without the 
effects of tensor coupling. 
So we conclude that the vector meson-nucleon 
 tensor coupling as well as pions play an
important role in decreasing the $g_{\rho NN}$ with density.

\begin{figure}
\vskip -1.5cm
\centerline{\epsfig{file=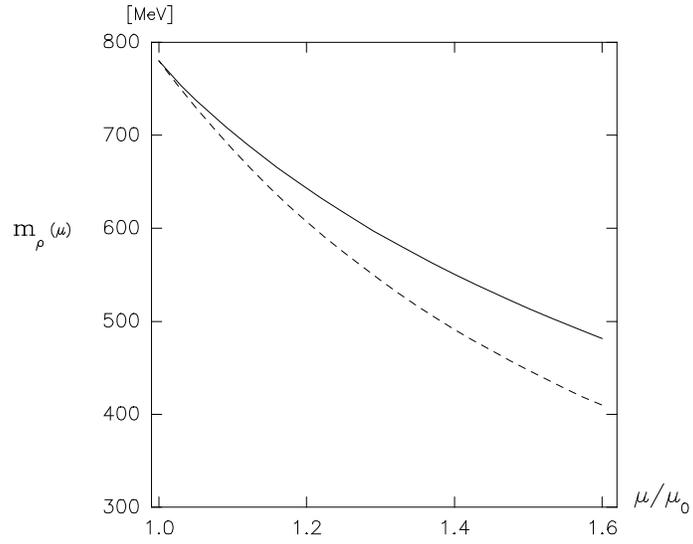,width=9cm}}
\caption{\small In-medium $\rho$-meson mass. 
 Solid line represent the in-medium $\rho$-meson mass with 
density dependent coupling $g^2(\mu)$
 and dashed lines represent the case of density independent 
coupling $g^2$. }\label{fig5}
\end{figure} 

\section{Summary and Discussion}
To study the properties of $\rho$-meson at finite density,  we construct
 the effective lagrangian in which $\rho$-meson is identified
as a dynamical gauge boson of hidden local symmetry.
We have calculated
the $\beta$-function for $\rho$-nucleon-nucleon coupling constant
($g_{\rho NN}$) and $\gamma$-function for $\rho$-meson mass
($\gamma_{m_\rho}$) at one loop approximation 
in the low energy limit. 

\begin{figure}
\centerline{\epsfig{file=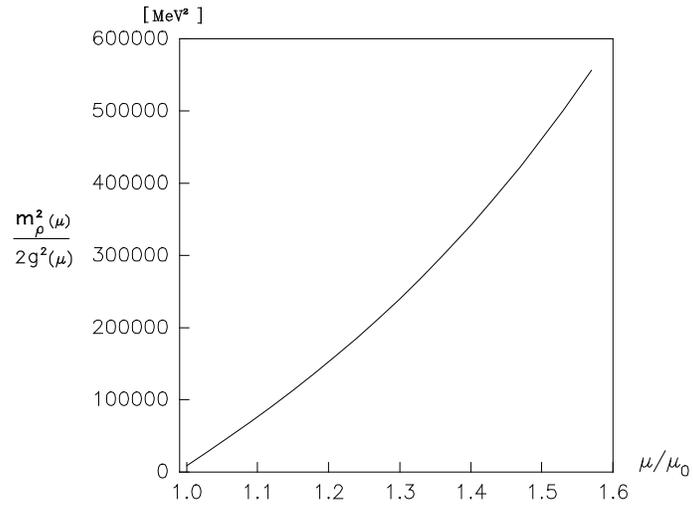,width=9cm}}
\caption{\small The vertical axis denotes the ratio of in-medium $m_\rho^2$ to
in-medium $2g^2$. }\label{fig6}
\end{figure} 

It is shown
that these $\beta$-function and $\gamma$-function calculated in free space
can be also used as  $\beta$-function and $\gamma$-function in hadronic
medium after changing the renormalization scale by the chemical potential
as far as the thermodynamical potential is concerned.   We also explicitly
demonstrate  that the $\rho$-meson mass and
coupling constant for $\rho$-nucleon-nucleon vertex drop as density increases.
Especially, we show that the pions and vector meson-nucleon 
 tensor coupling give a dominant
contribution to the $\beta$-function for $g_{\rho NN}$ and find that
$g_{\rho NN}$ drops significantly 
even in normal nuclear matter density$(\rho_0)$.
So the density dependence of the coupling constants
can affect the density dependence of $\rho$-meson mass substantially
even in normal nuclear matter density
because the $\rho$-meson self energy is a function of the coupling
constants.
For example, we solve eq.(\ref{rg22}) with density dependent coupling
constant($g^2(\mu)$) given in eq.(\ref{re2}). In Fig. \ref{fig5}, 
we depict in-medium
$\rho$-meson mass with density independent coupling constant(dashed line) and
the one with density dependent coupling constant(solid line). Although  
$g_{\rho NN}$ drops significantly 
even in normal nuclear matter density, the effects on in-medium $\rho$-meson
mass is not so significant. This is because the coefficient ($\frac{1}{2}
\frac{3}{(4\pi)^2}\sim 0.01 $)
of $g^2(\mu)$ in eq.(\ref{rg22}) is much smaller than $1$ and so the
change of $g^2(\mu)$ cannot modify the in-medium $\rho$-meson mass drastically.

Now we check whether the Kawarabayashi-Suzuki-Riazuddin-Fayyauddin
(KSRF) relation holds in medium or not. At zero-temperature and zero-density
the KSRF relation(for $\rho$-meson) is
\be
m_\rho^2=2g_{\rho\pi\pi}^2f_\pi^2.
\ee
It is easy to see that $g_{\rho NN}=g=g_{\rho\pi\pi}$ when $\kappa\rightarrow
0$ and $a=2$ from eq.(\ref{defi}).
To check whether the KSRF relation holds in medium or not, we plot
the quantity $m_\rho^2(\mu)/(2g^2(\mu))$ which
will be equal to $f_\pi^2(\mu)$
 if the KSRF relation holds in medium. 
In Fig. \ref{fig6}, we see that $f_\pi^2(\mu)$ increases with density.
But the $f_\pi^2(\mu)$ should decrease in medium\cite{br}\cite{kim}\cite{hara}.
Therefore, it seems that the KSRF relation does not hold in medium in our
calculations. 

In this work, we use several assumptions which are physically relevant for
our analysis to solve the renormalization group equations.
For the complete analysis, 
one may calculate, for example,  all the $\beta$-function and 
$\gamma$-functions and solve
coupled renormalization group equation as discussed in section 4.
However we do not expect any substantial change in our conclusion.  
It will be interesting to study the effects of resonances\cite{fp} 
and to see how the resonances affect our results.

Finally, we discuss whether the non-renormalizability of our effective
chiral lagrangian spoils the renormalization group arguments in section 3
or not.
As it is well known, we can get rid of all the infinites in a non-renormalizable
theory by including infinite number of counter terms allowed by symmetry
\cite{lecs}.
Calculating the thermodynamic potential(pressure) with a diagram 
which involves a vertex from a non-renormalizable interaction, 
we may get a coefficient which depends both on the
cut-off scale and on the chemical potential.
As in Refs.\cite{moki}\cite{mc2}, we do renormalizations
before we do the four momentum integrations which introduce density effects
 to our thermodynamic potential or pressure.
So it is impossible to get density or chemical potential($\mu$) dependent
coefficients during renormalization.
It is shown in the lectures\cite{lecs} that in a mass-independent renormalization scheme, 
loop integrals do not have a power law dependence on any big scale such as cutoff. 
In the case of a mass-independent renormalization scheme, cutoff or
renormalizaton scale only appears in logarithms.
Since we have used mass-independent renormalization scheme in this paper, 
we don't have any coeffient depending
on some power of cutoff or renormalization scale during renormalization.
 So we don't have a coefficient which depends both on the
cut-off scale and on the chemical potential except logarithmic dependence.
In Appendix, it is shown that this logarithmic dependence 
 plays a very important role in our renormalization group analysis
and therefore the non-renormalizability of our effective
chiral lagrangian might not be crucial for the renormalization group 
arguments in section 3.

\section*{Acknowledgments}
We would like to thank Gerry Brown and Mannque Rho for their invaluable
suggestions and  enlightening discussions. This work is
 supported in part by the Korean Ministry of
Education(BSRI 98-2441) and in part by
KOSEF(Grant No. 985-0200-001-2).

\setcounter{equation}{1}
\renewcommand{\theequation}{A.\arabic{equation}}
\section*{Appendix}

In this Appendix, we present a simple model calculation to 
 show how to calculate thermodynamic potential
(pressure) and to see how the renormalization group equation
works which we set up with some dimensional 
coupling constants such as $\frac{\lambda}{f_\pi}$.
The lagrangian we are considering is given by
\ba
{\it L}=\frac{1}{2}\partial_\mu \pi \partial^\mu \pi
+ \bar\Psi (x) i\gamma^\mu \partial_\mu\Psi (x)
       -m\bar\Psi (x)\Psi (x)
    +\frac{\lambda}{f_\pi}\bar\Psi (x)
    \gamma_5\gamma^\mu \partial_\mu\pi\Psi (x). 
\ea
\begin{figure}[htb]
\centerline{\epsfig{file=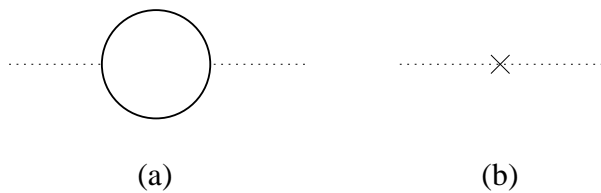,width=8cm}}
\caption{\small Self energy diagram for pion with counter term.
Dotted line is for pion and solid line represents fermion.}\label{af1}
\end{figure}
Since here we are not interested in the thermodynamic potential (pressure) itself,
we will consider only one type of graph and keep the terms which are essential in
our renormalization group analysis. It can be easily seen in the QED example in section
 3 that only the terms containing $\ln (\frac{\mu^2}{\Lambda^2})$ contribute to lowest order
$\beta$-function.
We also take the limit $\mu >> m$\cite{moki}\cite{mc2}.

\begin{figure}[htb]
\centerline{\epsfig{file=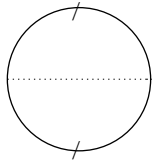,width=2cm}}
\caption{\small $2nd$ order contribution to the thermodynamic potential. Here solid lines with
$/$ denote the thermal(density) part of the fermion propagator\cite{lands}. }\label{af2}
\end{figure}

\begin{figure}[htb]
\centerline{\epsfig{file=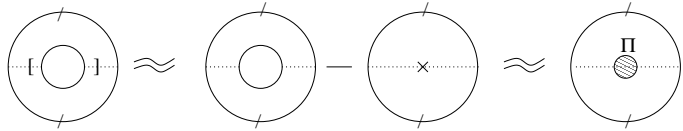,width=9cm}}
\caption{\small  $4th$ order contribution to the thermodynamic potential.
The brackets, $[$ $]$ indicate that we renormalize the enclosed diagram (pion self-energy)
and $\Pi$ represents the finite part of the pion self-energy. }\label{af3}
\end{figure}

Now consider Fig. \ref{af1} (a) which is calculated to be 
\ba
i\Pi (k^2)=\frac{(mf_\lambda)^2}{2\pi^2}k^2 i[\frac{1}{\epsilon} -\ln{\frac{k^2}{\Lambda^2}}]
\ea
where $f_\lambda\equiv \frac{\lambda}{f_\pi}$ and $\Lambda$ is a renormalization scale.
Introducing counter term, Fig. \ref{af1} (b), we obtain the 
pion wavefunction renormalization constant $Z_\pi$. Then,
we can define
\ba
f_\lambda^B=f_\lambda Z_\pi^{-1/2}\Lambda^\epsilon
\ea
and we get the $\beta$-function for $f_\lambda$,
\ba
\beta_\lambda =\frac{(mf_\lambda)^2}{2\pi^2}f_\lambda.\label{ae1}
\ea
If our renormalization group anlysis is correct, the $\beta$-function
 eq.(\ref{ae1}) must have something to do with a $\beta$-function defined
in medium. 
To see this, 
we calculate the thermodynmic potential(or pressure) from Fig. \ref{af2} and  Fig. \ref{af3}.

In  Fig. \ref{af3}, we schematically draw how to renormalize 
 the thermodynamic potential (pressure), for rigorous discussion see \cite{moki}\cite{mc2}.
The thermodynamic potential (pressure) from Fig. \ref{af2} and  Fig. \ref{af3} is given by
\ba
\Omega_R =
\frac{(mf_\lambda)^2}{16\pi^4}\mu^4 + 
\frac{(mf_\lambda)^4}{32\pi^6}\mu^4\ln \frac{\mu^2}{\Lambda^2}+ ...\label{ae2}
\ea
where the $...$ represents the terms which are not important when we discuss the 
renormalization group analysis. 
This is obvious when we take a look at QED $\beta$-function discussed in section 3.

Now we construct renormalization equation following the procedure in section 3.
Note that here we renormalize pion wavefunction only and therefore $f_\lambda$.
Then the renormalization group equation for the thermodynamic potential, in this simple case, is
given by
\ba
[\Lambda\frac{\partial}{\partial \Lambda} 
 +\beta_{f_\lambda^R} (mf_\lambda^R)
\frac{\partial}{\partial f_\lambda^R}]
\Omega_R ( mf_\lambda^R, \mu, \Lambda) =0\label{rga}
\ea
where we use $mf_\lambda^R$ instead of $f_\lambda^R$ 
because the coupling $f_\lambda^R$
carries a derivative(mass parameter) which is
$m$ in our one loop calculations and  the subscript $R$ denotes
the renormalized parameters.
The identity based on dimensional analysis is found to be
\ba
[\Lambda\frac{\partial}{\partial \Lambda}+
\mu\frac{\partial}{\partial \mu} -4]
\Omega_R ( mf_\lambda^R,\mu,\Lambda)=0\label{ida}.
\ea
From eq.(\ref{rga}) and eq.(\ref{ida}), we can write  
the following renormalization group equation
\ba
[\mu\frac{\partial}{\partial \mu}  -\beta_{f_\lambda} (mf_\lambda^R)
\frac{\partial}{\partial f_\lambda^R} -4]
\Omega_R (mf_\lambda^R, \mu, \Lambda) =0\label{rgfa}.
\ea
The general solution is given by
\be
\Omega_R (mf_\lambda^R, \mu, \Lambda)=\mu^4
{\bar \Omega}_R
( mf_\lambda(\mu),\mu , \Lambda)\label{sola}
\ee
with  a density dependent(effective or running)
coupling constants defined by
\ba
\mu\frac{d}{d \mu}f_\lambda(\mu)&=& \beta_\lambda(
mf_\lambda(\mu)).\label{aa}
\ea

Using the explicit form of the thermodynamic potential of eq. (\ref{ae2}),
we can show explicitly that  
 the renormalization group equation, eq.(\ref{rgfa}),
can be satisfied  by identifyng $\beta_\lambda$ in eq. (\ref{aa}) with 
the one in eq.(\ref{ae1})  to order of $m^4f_\lambda^4$, 
\ba
&&[\mu\frac{\partial}{\partial\mu} - \frac{m^2f_\lambda^2(\mu)}{2\pi^2}f_\lambda (\mu)
 \frac{\partial}{\partial f_\lambda}  -4]\no
&&\cdot (\frac{m^2f_\lambda^2(\mu)}{16\pi^4}\mu^4 + 
\frac{m^4f_\lambda^4(\mu)}{32\pi^6}\mu^4\ln \frac{\mu^2}{\Lambda^2}+ ... )
=0.
\ea


\newpage
\begin{thebibliography}{99}
\bibitem{rho} G.E. Brown, G.Q. Li, R. Rapp, M. Rho
    and J. Wambach, Acta. Phys. Pol. {\bf B29}, 2309 (1998);
Y. Kim, R. Rapp, G.E. Brown and Mannque Rho, ``A Schematic Model for
Density Dependent Vector Meson Masses,'' nucl-th/9902009
\bibitem{friman} B. Friman, in Proceedings of the APCTP Workshop on
'Hadron Properties in Medium', Seoul, Korea, Oct. 27-31, 1997, to be 
published.
\bibitem{klingl} F. Klingl, N.Kaiser and W. Weise, 
            Nucl. Phys. {\bf A 624}, 527 (1997)

\bibitem{ceres} G. Agakichiev et al., Phys. Rev. Lett. {\bf 75}, 1272 (1995)
\bibitem{br} G.E. Brown and M. Rho, Phys. Rev. Lett. {\bf 66}, 2720 (1991)
\bibitem{kim}Y. Kim, H. K. Lee and M. Rho, Phys. Rev. {\bf C52}, R1184 (1995);
           Y. Kim and H. K. Lee, Phys. Rev. {\bf C55}, 3100 (1997)
\bibitem{roy} A. K. D. Mazumder, B. D. Roy, A. Kundu and T. De,  
Phys. Rev. {\bf C53}, 790 (1996)
\bibitem{rapp} R. Rapp, G. Chanfray and J. Wambach,
Nucl. Phys. {\bf A 617}, 472 (1997)
\bibitem{kugo} M. Bando, T. Kugo and K. Yamawaki, Phys. Rep. {\bf 164}, 217 
(1988) ; M. Bando et al., Phys. Rev. Lett. {\bf 54}, 1215 (1985)
\bibitem{ka} J. I. Kapusta, Nucl. Phys. {\bf B148}, 461 (1979)
\bibitem{ma} M. Matsumoto, Y. Nakano and H. Umezawa,
            Phys. Rev. {\bf D29}, 1116(1984)
\bibitem{cp} J.C. Collins and M.J. Perry, 
  Phys. Rev. Lett {\bf 34}, 1353 (1975)
\bibitem{moki}  P.D. Morley and M.B. Kislinger, Phys. Rep. {\bf 51}, 63 (1979)
\bibitem{mc2} B.A. Freedman and L. D. McLerran, Phys. Rev. {\bf D16},
              1147 (1977)
\bibitem{mc} B.A. Freedman and L. D. McLerran, Phys. Rev. {\bf D16},
              1169 (1977)
\bibitem{furi} S. Furui, R. Kobayashi and M. Nakagawa 
 Nuovo Cimento {\bf 108A}, 241 (1995) 
\bibitem{hara} M. Harada and A. Shibata, Phys. Rev. {\bf D55}, 6716 (1997)
\bibitem{bor} G.E. Brown, E. Osnes and M. Rho, Phys. Lett. {\bf 163}, 41 (1985)
\bibitem{bon} R. Machleidt, K. Holinde and C. Elster, Phys. Rep. {\bf 149},
 1 (1987)
\bibitem{ha}M. Harada and K. Yamawaki, Phys. Lett. {\bf B297}, 151(1992)
\bibitem{bla}J.-P. Blaizot, J. Korean Phys. Soc. {\bf 25}, S65 (1992)
\bibitem{eric}Y. Ericson and W. Weise, {\it Pions and Nuclei} (Oxford
Science Publications, 1988).
\bibitem{gasser} J. Gasser, M.E. Sainio and A. Svarc, Nucl. Phys.
{\bf B307}, 779 (1988)
\bibitem{rhop} M. Rho, Phys. Rept. {\bf 240}, 1 (1994)
\bibitem{song} C. Song, G.E. Brown D. -P. Min and M. Rho, 
     Phys. Rev. {\bf C56}, 2244 (1997)
\bibitem{rho88} M. Rho, Phys. Rev. Lett. {\bf 54}, 767 (1985)
\bibitem{qu} M. Quiros, ``Constraints on the higgs boson properties
from the effective potential,'' hep-ph/9703412
\bibitem{fp} B. Friman and H.J. Pirner, Nucl. Phys.
{\bf A 617}, 496 (1997); W. Peters, M. Post, H. Lenske, S. Leupold 
and U. Mosel, Nucl. Phys. {\bf A 632}, 109 (1998)
\bibitem{lecs}D. B. Kaplan, ``Effective Field Theories,'' nucl-th/9506035;
A. V. Manohar, ``Effective Field Theories,'' hep-ph/9606222;
A. Pich, ``Effective Field Theory,'' hep-ph/9806303
\bibitem{lands}N.P. Landsman and C.G. van Weert, Phys. Rept. {\bf 145}, 141 (1987) 
\end{thebibliography}
\end{document}